\documentstyle[12pt]{article}
\let\ssection=\section
\renewcommand{\section}{\setcounter{equation}{0}\ssection}

\newcommand{\be}{\begin{enumerate}}
\newcommand{\ee}{\end{enumerate}}
\newcommand{\bi}{\begin{itemize}}
\newcommand{\ei}{\end{itemize}}
\newcommand{\beq}{\begin{equation}}
\newcommand{\eeq}{\end{equation}}
\newcommand{\beqa}{\begin{eqnarray}}
\newcommand{\eeqa}{\end{eqnarray}}

\newcommand{\eq}[1]{(\ref{#1})}
\newcommand{\eqs}[2]{(\ref{#1}--\ref{#2})}
\newcommand{\ie}{{\em i.e.,\ }}
\newcommand{\gmh}{Gell-Mann and Hartle}

\renewcommand{\a}{\alpha}                   
\renewcommand{\b}{\beta}                    
\renewcommand{\d}{\delta}					
\newcommand{\e}{\epsilon}

\newcommand{\g}{\gamma}
\newcommand{\k}{\kappa}
\renewcommand{\l}{\lambda}
\newcommand{\m}{\mu}
\newcommand{\n}{\nu}
\newcommand{\r}{\rho}

\newcommand{\bra}[1]{\langle{#1}|}
\newcommand{\braket}[2]{\bra{#1}{#2}\rangle}
\newcommand{\eqdef}{\stackrel{\rm def}{=}}
\newcommand{\ket}[1]{|#1\rangle}
\newcommand{\ketbra}[2]{\ket{#1}\bra{#2}}

\newcommand{\ebra}[2]{\langle \exp #1 (#2) |}
\newcommand{\ebraket}[4]{\ebra #1 #2 \exp #3 (#4) \rangle}
\newcommand{\eket}[2]{|\exp #1 (#2) \rangle}

\newcommand{\ot}{\otimes}

\newcommand{\tr}{{\rm tr}}
\newcommand{\half}{\frac{1}{2}}
\newcommand{\inner}[2]{\langle #1 ,#2\rangle}

\newcommand\mathC{\mkern1mu\raise2.2pt\hbox{$\scriptscriptstyle|$}
                {\mkern-7mu\rm C}}
\newcommand{\mathR}{{\rm I\! R}}                

\renewcommand\[{[\,}                            
\renewcommand\]{\,]}                            

\newcommand{\C}{{\tilde{C}}}
\newcommand{\D}{{\cal D}}

\renewcommand{\H}{{\cal H}}
\newcommand{\K}{{\cal K}}
\renewcommand{\L}{{\cal L}}
\renewcommand{\P}{{\cal P}}
\renewcommand{\S}{{\cal S}}
\newcommand{\Scts}{{S_{\rm cts}}}
\newcommand{\U}{{\cal U}}
\newcommand{\UP}{{\cal UP}}
\newcommand{\V}{{\cal V}}
\newcommand{\Vcts}{{\V_{\rm cts}}}

\newcommand{\lbra}{\left(}
\newcommand{\rbra}{\right)}

\newcommand{\hh}[2]{{#1}_{t_1},{#1}_{t_2},\ldots,{#1}_{t_#2}}

\newcommand{\nn}{\nonumber}

\begin{document}

\begin{titlepage}
\hspace{8truecm} Imperial/TP/94--95/19

\hspace{8truecm} DAMTP 95--16

\begin{center}
        {\large\bf Continuous Histories and the History Group in\\[0.2cm]
                  Generalised Quantum Theory}
\end{center}
\vspace{1 truecm}
\begin{center}
        C.J.~Isham\footnote{email: c.isham@ic.ac.uk}\\[0.4cm]
       	Blackett Laboratory\\
		Imperial College\\
		South Kensington\\
        London SW7 2BZ\\
        United Kingdom\\
\medskip
\begin{center} and \end{center}
\medskip
        N.~Linden\footnote{email: n.linden@newton.cam.ac.uk}\\[0.4cm]
        D.A.M.T.P.\\
        University of Cambridge\\
        Cambridge CB3 9EW\\
        United Kingdom\\
\end{center}

\begin{center} March 1995\end{center}

\begin{abstract}
We treat continuous histories within the histories approach to generalised
quantum mechanics. The essential tool is the `history group': the analogue,
within the generalised history scheme, of the canonical group of
single-time quantum mechanics.
\end{abstract}

\end{titlepage}
\section{Introduction}
In this paper we wish to illustrate and develop our algebraic scheme
\cite{Isham94a,IL94a,IL94b} for the consistent-histories
approach to quantum theory by extending it to include histories that
are {\em continuous\/} (rather than discrete) functions of time.  We
work within the approach to generalised quantum theory pioneered by
Griffiths \cite{Gri84}, Omn\`es
\cite{Omn88a,Omn88b,Omn88c,Omn89,Omn90,Omn92} and Gell-Mann and Hartle
\cite{GH90a,GH90b,GH90c,Har91a,Har91b,GH92,Har93a} which starts
from an observation in conventional quantum theory concerning the
joint probability of finding each of a time-ordered sequence of
properties $\a=\hh\a n$ with $t_1<t_2<\cdots<t_n$ (we shall call a
sequence of this type a {\em homogeneous history\/}, and refer to
the sequence of times as the {\em temporal support\/} of the
history).  Namely, if the initial state at time $t_0$ is a density
matrix $\r_{t_0}$ then the joint probability of finding all the
properties in an appropriate sequence of measurements is
\beq
 	{\rm Prob}(\hh\a n;\r_{t_0})=\tr_\H(\tilde C_\a^\dagger\r_{t_0}
			\tilde C_\a)  						\label{Prob:a1-an}
\eeq
where the `class' operator $ \tilde C_\a$ is given in terms of the
Schr\"odinger-picture projection operators $\a_{t_i}$ as
\beq
 	\C_\a:=U(t_0,t_1)\a_{t_1} U(t_1,t_2)\a_{t_2}\ldots U(t_{n-1},t_n)
        \a_{t_n}U(t_n,t_0)                		\label{Def:C_a}
\eeq
and where $U(t,t')=e^{-i(t-t')H/\hbar}$ is the unitary
time-evolution operator from time $t$ to $t'$. Note that our
operator $\C_\a$ is the adjoint of the operator $C_\a$ used by
Gell-Mann and Hartle.

    The main assumption of the consistent-histories interpretation of
quantum theory is that, under appropriate conditions, the probability
assignment \eq{Prob:a1-an} is still meaningful for a {\em closed\/}
system, with no external observers or associated measurement-induced
state-vector reductions (thus signalling a move from `observables' to
`beables'). The satisfaction or otherwise of these conditions (the
`consistency' of a complete set of histories: see below) is determined
by the behaviour of the {\em decoherence functional\/}
$d_{(H,\rho)}$. This is the complex-valued function of pairs of
homogeneous histories $\a=(\hh{\a}n)$ and
$\b=(\b_{t_1'},\b_{t_2'},\ldots,\b_{t_m'})$ defined as
\beq
    d_{(H,\rho)}(\a,\b)=\tr(\C_\a^\dagger\rho \C_\b)         \label{Def:d}
\eeq
where the temporal supports of $\a$ and $\b$ need not be the same.
Note that, as suggested by the notation $d_{(H,\rho)}$, both the
initial state and the dynamical structure (\ie the Hamiltonian $H$)
are coded in the decoherence functional. In our approach, a history
of the type $(\hh{\a}n)$ is just a `passive', time-ordered sequence
of propositions. Thinking of the Schr\"odinger-picture projectors
$\a_{t_1},\a_{t_2},\ldots$ as representing propositions, the
homogeneous history $(\hh{\a}n)$ can read as the sequential
proposition ``$\a_{t_1}$ is true at time $t_1$, and then $\a_{t_2}$
is true at time $t_2$, and then \ldots, and then $\a_{t_n}$ is true at
time $t_n$''.

	An important suggestion by \gmh\ is that, in the generalised
theory, a history should be regarded as a fundamental entity in its
own right, not necessarily just a time-ordered sequence of
projection operators.  The physical results are obtained by
calculating the decoherence functional, now defined as a
complex-valued function of pairs of histories that satisfies certain
algebraic conditions. When applied to standard quantum theory, a
significant technical supposition is that the class of histories
should be extended to include `inhomogeneous' histories,
\ie propositions obtained by applying the logical `or' operation to
homogeneous histories (for example, ``$\a_{t_1}$ is true at time
$t_1$, and then $\a_{t_2}$ is true at time $t_2$, {\em or\/},
$\b_{t_1'}$ is true at time $t_1'$, and then $\b_{t_2'}$ is true at
time $t_2'$''.

	In previous papers \cite{Isham94a,IL94a,IL94b} we have argued
that the basic ingredients of this generalised quantum theory should
be viewed as the set of histories $\UP$ (or, more accurately, the set
of {\em propositions about\/} histories) and the set of decoherence
functionals $\D$, with the pair ($\UP,\D$) being regarded as the
analogue in the history theory of the pair ($\L,\S$) in standard
quantum theory, where $\L$ is the lattice of propositions and $\S$
the space of states on $\L$.

	As explained in \cite{Isham94a,IL94a}, there are cogent reasons
for postulating that the natural mathematical structure on the set
of histories, $\UP$, is that of an {\em orthoalgebra\/}
\cite{FGR92}, with the three orthoalgebra operations $\oplus,\neg$
and $<$ corresponding respectively to the disjoint sum, negation and
coarse-graining operations invoked by \gmh. The properties of the
decoherence functional $d: \UP
\times \UP \rightarrow
\mathC$ are
\be
    \item {\em Hermiticity\/}: $d(\a,\b)=d(\b,\a)^*$ for all
          $\a,\b\in\UP$.
    \item {\em Positivity\/}: $d(\a,\a)\ge0$ for all $\a\in\UP$.
    \item {\em Additivity\/}: if $\a$ and $\b$ are disjoint then, for
          all $\g$, $d(\a\oplus\b,\g)=d(\a,\g)+d(\b,\g)$. If
          appropriate, this can be extended to countable sums.
    \item {\em Normalisation\/}: $d(1,1)=1$.
\ee

	One important motivation for our framework is the fact that
discrete-time histories in quantum mechanics can indeed be given the
structure of an orthoalgebra. The key idea is that an $n$-time,
homogeneous history proposition $(\hh\a n)$ can be associated with
the operator $\a_{t_1}\ot\a_{t_2}\cdots\ot\a_{t_n}$ which is a genuine {\em
projection\/} operator on the $n$-fold tensor product
$\H_{t_1}\ot\H_{t_2}\ot\cdots\ot\H_{t_n}$ of $n$-copies of the
Hilbert-space $\H$ on which the canonical theory is defined
\cite{Isham94a,IL94a}.

	In this paper we investigate how {\em continuous\/}-time history
propositions in standard quantum mechanics can be treated within
this framework.  This raises two immediate issues, both of which
we address:
\be

\item the construction of a continuous analogue of the finite
{\em product\/} of projection operators used in the definition of
the class operator in \eq{Def:C_a};

\item the construction of a continuous analogue of the finite {\em
tensor product\/} of projection operators used to associate a
homogeneous history with a projection operator.
\ee

	We shall approach this task with the aid of a tool that we
believe is of considerable interest and importance in its own right.
This is a {\em history group\/}: an analogue for the history theory
of the canonical group used in single-time quantum mechanics. The
key idea is that a unitary representation of the appropriate history
group leads naturally to an orthoalgebra of projection operators
that are to be interpreted as propositions about the `histories' of
that theory. These projection operators are the elements of the
spectral representations of the self-adjoint generators of the Lie
algebra of the group, or other operators closely related to them.
In the case of standard, but continuous-time, quantum theory we seek
a history group whose representations yield projection operators
that can be associated with propositions about continuous-time
histories.

	We will demonstrate the existence of such a group for the
standard quantum mechanics of a particle moving on the real line,
and also construct the decoherence functionals that permit the
potential assignment of probabilities to suitably coarse-grained
histories for this system.  As in the case of discrete-time
histories \cite{IL94b}, we show that the decoherence functional may
be written on a tensor product space in terms of an operator $X$
that carries all the information about dynamics. This emphasises the
fact that, within the history approach, dynamical evolution is
described by a single operator rather than the usual one-parameter
family of unitary operators.

\section{Kinematics}
\subsection{The History Group}
As motivation for what follows, let us consider $n$-time quantum
mechanics of a particle moving on the line $\mathR$.  As explained in
the Introduction, a homogeneous history $\a = (\hh\a n)$ can be
associated with a projection operator $\a_{t_1}\ot\a_{t_2}\cdots\ot\a_{t_n}$ on
the $n$-fold tensor product $\V_n = \H_{t_1}\ot\H_{t_2}\cdots\H_{t_n}$
of $n$-copies of the Hilbert-space $\H$ of the canonical theory.
Although we have not made direct use of it in our work so far, it is
clear that, since $\H$ carries a representation of the Heisenberg-Weyl
group with Lie algebra (we set $\hbar=1$)
\beq
	\[x, p\] = i,
\eeq
the Hilbert space $\V_n$ carries a unitary representation of the
$n$-fold product group with Lie algebra generators satisfy
\beqa
	\[x_k,x_m\]&=& 0		\label{discreteHWxx}	\\
	\[p_k,p_m\]&=& 0		\label{discreteHWpp}	\\
	\[x_k,p_m\]&=& i\d_{km}	\label{discreteHWxp}
\eeqa
with $k,m = 1,2,\ldots,n$.  Thus, although the vectors of $\V_n$ are
not directly related to decoherence functionals (which can be viewed
as the history analogues of states in single-time quantum theory),
it is clear that the Hilbert space $\V_n$ carries a representation
of the `history group' with Lie algebra
\eqs{discreteHWxx}{discreteHWxp}.  In fact, we could turn the
discussion around and {\em define\/} the history version of $n$-time
quantum mechanics by starting with \eqs{discreteHWxx}{discreteHWxp}. In
this approach, $\V_n$ arises as a representation space for
\eqs{discreteHWxx}{discreteHWxp}, and then any tensor products
$\a_{t_1}\ot\a_{t_2}\ot\cdots\ot\a_{t_n}$ corresponding to
sequential histories of values of position or momentum (or linear
combinations of them) are indeed elements of the spectral
representations of this Lie algebra.

	This is the approach that we have found fruitful for
discussing continuous-time histories.  Specifically, motivated by
\eqs{discreteHWxx}{discreteHWxp}, we start with the
history-group with the following Lie algebra
\beqa
	\[x_{t_1},x_{t_2}\]&=&0				\label{ctsHWxx}	\\
	\[p_{t_1},p_{t_2}\]&=&0				\label{ctsHWpp}	\\
	\[x_{t_1},p_{t_2}\]&=&i\d(t_1-t_2)	\label{ctsHWxp}
\eeqa
where $-\infty\leq t_1,\,t_2\leq\infty$.  The operators appearing
here should {\em not\/} be confused with the family $x(t), p(t)$ of
Heisenberg-picture operators in normal quantum theory. Indeed,
it should be noted that, although we are studying the history
version of quantum {\em mechanics\/} on $\mathR$, we are led to a
Lie-algebra which usually arises as the canonical commutation
relations of a one-dimensional quantum field theory!
In what follows, we shall need
to be a little more careful about how this algebra is specified; in
particular, we introduce a test-function space which we take
to be the space $L_{\mathR}^2(\mathR)$ of real, square-integrable,
functions on $\mathR$.  Hence \eqs{ctsHWxx}{ctsHWxp} are replaced by
\beqa
	\[x_f,x_g\]&=&0						\label{SmearedCtsHWxx}	\\
	\[p_f,p_g\]&=&0						\label{SmearedCtsHWpp}	\\
	\[x_f,p_g\]&=&i(f,g) 				\label{SmearedCtsHWxp}
\eeqa
where $f,g\in L^2_{\mathR}(\mathR)$ and
$(f,g)\eqdef\int_{-\infty}^\infty f(x)g(x)\,dx$.

\subsection{The Hilbert Space $\Vcts$}
In order to understand how the representations of
\eqs{ctsHWxx}{ctsHWxp} lead to an appropriate notion of a
continuous tensor product of copies of the Hilbert space
$L^2(\mathR)$ of canonical quantum mechanics, we need to introduce
the concept of an exponential Hilbert space
\cite{ArakiWoods66,Klau69,Streater69,Guich72,Erven81}.  Given a
Hilbert space $\K$, the exponential Hilbert space $e^\K$ is
constructed as follows.  Let $(\ot\K)^n$ denote the $n$-fold tensor
product of $\K$ with itself, and let $(\ot\K)^n_S$ be the subspace
of $(\ot\K)^n$ spanned by the vectors $(\ot\phi)^n\ ,\phi\in\K$,
where $(\ot\K)^0 = (\ot\K)^0_S$ is defined to be the one-dimensional
Hilbert space of the complex numbers.  Then the exponential Hilbert
space is defined as $e^\K\eqdef\oplus_{n=0}^\infty(\ot K)^n_S$.  Of
course, the space $e^\K$ is also known as the `bosonic Fock space'
over $\K$.  It is more usual to define $(\ot\K)^n_S$ as
the subspace of $(\ot\K)^n$ spanned by symmetrised product vectors;
however, the definition used here is equivalent (for example, in
$(\ot\K)^2$, one may obtain $a\ot b + b\ot a$ as $(a+b)\ot(a+b)-a\ot
a-b\ot b$.)

	Let $\ket{\exp\phi}$ denote the (non-normalised) coherent state vector
\beq
\oplus^\infty_{n=0} (n!)^{-{1\over 2}} \lbra \ot\ket\phi \rbra^n.
\eeq
Then the inner product in $e^\K$ is given by
\beq
		\braket{{\exp\phi}}{\exp\psi}_{e^\K} = e^{\inner\phi\psi_{\K}}.
\eeq
As shown in \cite{ArakiWoods66}, the vectors $\ket{\exp\phi}$ are total in
$e^\K$: a fact we shall use frequently later.

	Let $\cal T$ denote the complexification of the real test
function space used to define smeared canonical commutation
relations, so that we can write
\beqa
		\[a_f,a_g\] 				&=& 0		\label{CCRaa}	\\
		\[a_f^\dagger,a_g^\dagger\] &=& 0		\label{CCRadad}	\\
		\[a_f,a_g^\dagger\] 		&=& \inner{f}{g}_{\cal T}.
												\label{CCRaad}
\eeqa
where $\inner{f}{g}$ denotes the associated scalar product.
Then the Fock representation of these creation and annihilation
operators is defined on the exponential Hilbert space $e^{\cal T}$
so that, for example, the matrix element of $a_f$ is
\beq
	\bra{\exp\phi}a_f\ket{\exp\psi}_{e^{\cal T}} =
			\inner{f}\psi e^{\inner{\phi}{\psi}_{\cal T}}.
\eeq

	A special case is when $\cal T$ is $\mathC$, which we shall treat
explicitly here as the results are needed later. The usual
normalised coherent states are
\beq
	\ket z \eqdef e^{-\half |z|^2 +za^\dagger}\ket 0,
\eeq
and satisfy $\braket zw = e^{-\half |z|^2-\half |w|^2 +z^*w}$.  These
normalised states are related to the exponential vectors by
$\ket{\exp z}= e^{\half|z|^2}\ket{z}$, and the Hilbert
space $\exp\,{\mathC}$ is  isomorphic to $L^2(\mathR)$ via
\beqa
	\exp\,\mathC &\simeq& L^2(\mathR,dx)\label{ExpC}		\\
	\ket{\exp z} &\mapsto& \braket{x}{\exp z}=(2\pi)^{-{1\over 4}}
			e^{zx-\half z^2 -{1\over 4}x^2}.				\nn
\eeqa

	Our reason for introducing exponential Hilbert spaces is the
existence of a particularly convenient construction of a continuous
tensor product of a one-parameter family of them.  In general,
given a family $t\mapsto \H_t$ of Hilbert spaces, we
try to define an inner product, following Streater, \cite{Streater69}
as
\beq
{\braket{{\ot_t u_t}}{{\ot_t v_t}}}_{\ot_t\H_t} \eqdef
e^{\int_{-\infty}^\infty\log {\inner{{u_t}}{ v_t}}_{\H_t}\,dt}
\eeq
if this expression makes sense. This is intended to be the
continuous analogue of the inner product between discrete tensor
products of vectors
\beq
	\braket{u_1\ot u_2\ot\cdots\ot u_n}{v_1\ot v_2\ot\cdots\ot v_n}
	\eqdef \prod_{i=1}^n\inner{u_i}{v_i}\equiv
		e^{\sum_{i=1}^n\log\inner{u_i}{v_i}}.
\eeq
If $\H_t$ is an exponential Hilbert space $\H_t=e^{\K_t}$, then the
construction works since
\beq
{\braket{{\exp \phi_t}}{{\exp \psi_t}}}_{e^{\K_t}} {=}
e^{\inner{\phi_t}{\psi_t}_{\K_t}}
\eeq
and so the definition of the scalar product on the continuous tensor
product of copies of $e^{\K_t}$ as
\beq
{\braket{{\ot_t\exp \phi_t}}{{\ot_t\exp\psi_t}}}_{\ot_t e^{\K_t}}\eqdef
e^{\int_{-\infty}^\infty {\inner{{\phi_t}}{{\psi_t}}}_{\K_t}\,dt}
\eeq
is well-defined.

Furthermore, the scalar product ${\int_{-\infty}^\infty
{\inner{{\phi_t}}{{\psi_t}}}_{\K_t}\,dt}$
is the inner product on the direct-integral Hilbert space $\int^\oplus \K_t$,
and hence
\beq
{\braket{{\ot_t\exp \phi_t}}{{\ot_t\exp \psi_t}}}_{\ot_t e^{\K_t}} {=}
 {\ebraket{\phi}{\cdot}{\psi}{\cdot}}_{\exp\int^\oplus \K_t}
\eeq
In fact, there is an isomorphism
\beqa
	\ot_t \exp \K_t &\simeq& \exp\int^\oplus\K_t \label{ExpIso} \\
	\ot_t\ket{\exp\phi_t}&\mapsto& \ket{\exp\phi(\cdot)}			\nn
\eeqa

	Let us use these ideas now  to see how the Fock representation of
\eqs{SmearedCtsHWxx}{SmearedCtsHWxp} with the test-function space
$L_{\mathR}^2(\mathR)$ leads to a continuous tensor product space $\ot_t
L^2_t(\mathR)$.  The complexification of the real vector space
$L^2_\mathR(\mathR)$ is just $L^2(\mathR)$, and hence the Fock
representation of \eqs{CCRaa}{CCRaad} is on the
space $\exp(L^2(\mathR))$.  However, $L^2(\mathR)$ is isomorphic to
the direct integral $\int^\oplus \mathC_t\,dt$ via
\beqa
	\int^\oplus \mathC_t\,dt &\simeq& L^2(\mathR,dt) \\
	\int^\oplus w_t\,dt &\mapsto& w(\cdot)
\eeqa
so that we may think of the Hilbert space as
\beq
	\exp(L^2(\mathR))\simeq\exp\int^\oplus \mathC_t\,dt.
\eeq
On the other hand, \eq{ExpIso} means that
\beq
	\exp\int^\oplus \mathC_t\,dt \simeq \ot_t\lbra\exp\mathC_t\rbra,
\eeq
and \eq{ExpC} shows that
\beq
	\exp\,\mathC_t\simeq L^2_t(\mathR).
\eeq
Hence we see that the Hilbert space (\ie Fock space) upon which the
history algebra \eqs{SmearedCtsHWxx}{SmearedCtsHWxp} is naturally
represented, is isomorphic to
$\ot_t L^2_t(\mathR)$,  and therefore the space $\Vcts$ which
carries the propositions/projection operators of our continuous history
theory is
\beq
	\Vcts\eqdef\ot_t\lbra L^2_t(\mathR)\rbra
		\simeq \exp\lbra L^2(\mathR,dt)\rbra.
\eeq

\subsection{History Propositions}
Having found the Hilbert space $\Vcts$, the next step is to identify
some projection operators on $\Vcts$ that have a clear physical
meaning. In standard canonical quantum theory, the Lie algebra of
the canonical group provides a preferred class of classical
observables that are to be quantised as self-adjoint operators. A
typical proposition is of the form that the value of such an
observable lies in some specified range. This proposition is
represented mathematically by the appropriate projector in the
spectral representation of the associated self-adjoint operator.

	 It would be possible to develop the analogue of this idea in
the history theory. In particular, using the smeared form
\eqs{SmearedCtsHWxx}{SmearedCtsHWxp} of the history algebra, it
is natural to consider the quantum representation of propositions of
the form ``$x_f+p_g$ lies in a subset $\Delta\subset\mathR$'', where
$f$ and $g$ are test functions. Propositions of this type clearly
deal with the time {\em averages\/} of position and momentum.

	However, in this paper we shall adopt a somewhat different approach,
based on the observation that coherent states play an intimate
role in the construction of the continuous tensor product. This
suggests that it may be productive to focus on projectors onto such
states. Thus the task is to define a continuous tensor product
$\ot_t P_{\l(t)}$ where $t\mapsto \l(t)$ is a complex-valued map of
bounded variation from $\mathR$, and where, for each $t\in\mathR$,
the operator $P_{\l(t)}\eqdef\ket{\l(t)}\bra{\l(t)}$ is the
projector onto the normalised coherent state $\ket{\l(t)}$.

	The first obvious thing to try is
\beq
	(\ot_t P_{\l(t)})\left(\ot_t\ket{\exp\m(t)}\right) \eqdef
		\ot_t \left(P_{\l(t)} \ket{\exp\m(t)}\right) 		\label{badproja}
\eeq
where we recall that the exponential states $\ket{\exp\m(t)}\in
L^2_t(\mathR)$ are given in terms of the normalised coherent states
$\ket{\m(t)}$ as
\beq
	\ket{\exp\m(t)} = e^{\half|\m|^2}\ket{\m(t)},
\eeq
and with
\beq
	\braket{{\l(t)}}{{\m(t)}}=
		e^{\l(t)^*\m(t)-\half|\l(t)|^2-\half|\m(t)|^2}.
\eeq
Thus
\beqa
	P_{\l(t)}\ket{\exp\m(t)} &=&
			e^{-\half|\l(t)|^2 +\l(t)^*\m(t)}\ket{\l(t)}\\
 			&=& e^{-\l(t)^*(\l(t) -\m(t))} \ket{\exp\l(t)},
\eeqa
and hence
\beq
	\ot_t P_{\l(t)} \left( \ot_t \ket{\exp\m(t)}\right) =
	\ot_t\left( e^{-\l(t)^*(\l(t) -\m(t))}\ket{\exp\l(t)}\right).
													\label{badprojb}
 \eeq
However, if $\m(t)$ and $\l(t)$ differ from each other on an open subset
of $t$ values, then the continuous product
\beq
	\prod_t\, e^{-\l(t)^*(\l(t)-\m(t))}
\eeq
will not converge. Hence the right-hand side of \eq{badprojb} is
not defined, and so we cannot use \eq{badproja}.

 	The difficulty we have found arises from the fact that it is not
the continuous product of normalised states $\ot_t\ket{\l(t)}$ that
is well-defined but rather the continuous product of {\em
non\/}-normalised states $\ot_t\ket{\exp\l(t)} = \ot_t
e^{\half|\l(t)|^2}\ket{\l(t)}$.  This suggests that we identify
$\ot_t P_{\l (t)}$ with $ P_{\ot_t\ket{\exp\l(t)}}$, \ie the
projector onto the vector $\ot_t\ket{\exp\l(t)}$ in $\ot_t
L_t^2(\mathR)$.  Under the identification of $\ot_t L^2_t (\mathR)$
with $\exp L^2(\mathR,dt)$, this projector is identified with $
P_{\ket{\exp\l(\cdot)}}=
e^{-\inner\l\l}\ket{\exp\l(\cdot)}\bra{\exp\l(\cdot)}$.  The action
of this projector is
\beq
	P_{\ket{\exp\l(\cdot)}}\ket{\exp\m(\cdot)} =
		e^{\inner{\l}{\m-\l}}\ket{\exp\l(\cdot)}
\eeq
where
\beq
	\inner {\l}{\m-\l}  = \int_{-\infty}^\infty
		\l^*(t)\left(\m(t)-\l(t) \right)\,dt.
\eeq
This operator {\em is\/} well-defined, and may easily be seen to satisfy
\beq
	P_{\ket{\exp\l(\cdot)}} P_{\ket{\exp\l(\cdot)}} =
		P_{\ket{\exp\l(\cdot)}}\quad \hbox{and} \quad
		 P_{\ket{\exp\l(\cdot)}}^\dagger = P_{\ket{\exp\l(\cdot)}},
\eeq
as it should.

	In addition to the projectors $P_{\ket{\exp\l(\cdot)}}$, we might
wish to handle propositions about histories that involve only a
finite time interval $[a,b]$.  Of course, one possibility is to
start with the Hilbert space $L^2[a,b]$ rather than $L^2(\mathR)$.
However, what we want is a single history theory that can accommodate
{\em all\/} possible finite time intervals, not just one.

	Since the continuous product of copies of the single projector
$P_{\ket{\l(a)}}$ is just itself, one might be tempted to define
$P_{\ket{\exp\l(t)}}$ for all $t$ as above but then impose the
constraints
\beqa \l(t) &=& \l(a)\quad t<a, \nn					\\
            &=& \l(b)\quad t>b.
\eeqa
However, the fact that the function $\l$ is supposed to be a member
of $L^2(\mathR,dt)$ leads to a difficulty: unless $\l(a)=\l(b)=0$,
the function $\l$ defined above will not be square-integrable, and
so the class of projectors $\{P_{\ket{\exp\l(\cdot)}}\}$ does not
contain operators of this type.

	What we would really like to do is to construct projectors that are equal
to $P_{\ket{\exp\l(\cdot)}}$ in the `active' region $[a,b]$, and are equal
to the unit operator outside $[a,b]$. This leads us to define the
following operator $P^{[a,b]}_{\ket{\exp\l(\cdot)}}$:
\beq
	P^{[a,b]}_{\ket{\exp\l(\cdot)}} \ket{\exp\m(\cdot)}\eqdef
		\exp\left(\int_a^b \l^*(t)(\m(t)-\l(t))\,dt\right)
			\ket{\exp\l\star\m(\cdot)}					\label{Pabl}
\eeq
where
\beq
	(\l\star\m)(t)\eqdef\left\{ \begin{array}{ll}
                     \l (t)\quad \mbox{if $t\in [a,b]$},	\\
                     \m (t)\quad \mbox{otherwise}.
                     \end{array}
\right.
\eeq

	A little work is required to show that $P^{[a,b]}_{\ket{\exp\l(\cdot)}}$
is a genuine projection operator on $\Vcts = \exp\lbra
L^2(\mathR)\rbra$.  Firstly consider
\beqa
	P^{[a,b]}_{\eket\l\cdot}P^{[a,b]}_{\eket\l\cdot}\eket{\m}{\cdot}
	&=&	e^{\int_a^b \l^*(t)(\m(t)-\l(t))\,dt}\,
 			P^{[a,b]}_{\eket \l \cdot}\eket {\l\star\m}\cdot\nn			\\
	&=& e^{\int_a^b \l^*(t)(\m(t)-\l(t))\,dt}\,
		e^{\int_a^b \l^*(t)((\l\star\m)(t)-\l(t))\,dt}\nn	\\
	&{}&\quad\times\quad\eket{{\l\star(\l\star\m)}}\cdot .
\eeqa
Now, if $t\in [a,b]$, then $\l\star\m(t)=\l(t)$, and hence the
second exponent vanishes. Furthermore, if $t\in [a,b]$, then
$\l\star(\l\star\m)(t)=\l(t)$, and if $t\not\in [a,b]$, then
$\l\star(\l\star\m)(t)=\m(t)$; hence $\l\star(\l\star\m) =
(\l\star\m)$. Thus we find
\beqa
	P^{[a,b]}_{\eket\l\cdot} P^{[a,b]}_{\eket\l\cdot}\eket{\m}{\cdot}
		&=& e^{\int_a^b \l^*(t)(\m(t)-\l(t))\,dt}\,
						\eket{{(\l\star\m)}}\cdot \nn					\\
		&=& P^{[a,b]}_{\eket \l \cdot} \eket{\m}{\cdot}.
\eeqa
Since this is true for all $\eket{\m}{\cdot}$ (\ie for all elements of a
total set of vectors) this implies that
\beq
	P^{[a,b]}_{\eket\l\cdot} P^{[a,b]}_{\eket\l\cdot}=
			P^{[a,b]}_{\eket\l\cdot}.
\eeq

	The other property we need to prove is self-adjointness.  To
this end, consider the matrix elements
\beqa
	\ebra \m\cdot P^{[a,b]}_{\eket \l \cdot} \eket\n\cdot
		&=& e^{\int_a^b \l^*(t)(\n(t)-\l(t))\,dt}\,
			\ebraket{\m}{\cdot}{\l\star\n}{\cdot}			\nn	\\
		&=& e^{\int_a^b \l^*(t)(\n(t)-\l(t))\,dt}\,			\nn \\
		&& \quad\times\quad e^{\int_{-\infty}^\infty \m^*(t)\,
				\l\star\n(t)\,dt}.
\eeqa
Now
\beq
	\int_{-\infty}^\infty \m^*(t)\,\l\star\n(t)\,dt =
		\int_{-\infty}^\infty \m^*(t)\n(t)\,dt +
			\int_{a}^b\m^*(t)(\l(t)-\n(t))\,dt
\eeq
and so
\beq
	\ebra \m\cdot P^{[a,b]}_{\eket \l \cdot} \eket\n\cdot
		=e^{\int_a^b(\m^*(t)-\l^*(t))(\l(t)-\n(t))\,dt}\,
			e^{\int_{-\infty}^\infty \m^*(t)\n(t)\,dt}.	\label{dum1}
\eeq
Now consider
\beqa
\ebra\m\cdot\lbra P^{[a,b]}_{\eket\l\cdot}\rbra^\dagger \eket\n\cdot
&=& \braket{{P^{[a,b]}_{\eket \l \cdot}\exp\m(\cdot)}}{{\exp\n(\cdot)}}\nn\\
&=&	e^{\int_a^b \l(t)(\m(t)-\l(t))^*\,dt}\,	\nn			\\
&{}&\quad\times\quad \ebraket{{\l\star\m}}{\cdot}{\n}{\cdot}	\nn			\\
&=& e^{\int_a^b \l(t)(\m(t)-\l(t))^*\,dt}	\nn			\\
&{}&\quad\times\quad
		e^{\int_{-\infty}^\infty(\l\star\m(t))^*\n(t)\,dt}.
\eeqa
But
\beq
	\int_{-\infty}^\infty (\l\star\m(t))^*\n(t)\,dt =
		\int_{-\infty}^\infty \m^*(t)\n(t)\,dt +
			\int_{a}^b(\l(t)-\m(t))^*\n(t)\,dt
\eeq
and so
\beq
	\ebra \m\cdot\lbra P^{[a,b]}_{\eket\l\cdot}\rbra^\dagger\eket\n\cdot
		= e^{ \int_a^b (\m^*(t)-\l^*(t))(\l(t)-\n(t))\,dt}
			e^{\int_{-\infty}^\infty\m^*(t)\n(t)\,dt}
\eeq
which is equal to \eq{dum1}. Thus the (bounded) operator
$ P^{[a,b]}_{\eket \l \cdot}$ satisfies
\beq
 	P^{[a,b]}_{\eket\l\cdot} = \lbra P^{[a,b]}_{\eket\l\cdot}\rbra^\dagger
\eeq
and
\beq
 	P^{[a,b]}_{\eket\l\cdot}P^{[a,b]}_{\eket\l\cdot}=P^{[a,b]}_{\eket\l\cdot}
\eeq
and is hence a projection operator on $\Vcts = \exp\lbra L^2(\mathR)\rbra$.

\section{Dynamics}
\subsection{The decoherence functional in terms of continuous products}
In our formulation of the \gmh\ generalised quantum theory, it is the
decoherence functional $d$ that contains all the information about the
dynamics.  In \cite{IL94b} it was shown that if $(\hh \a n)$ and
$(\b_{t_1^\prime},\b_{t_2^\prime},\ldots,\b_{t_{m}^\prime})$
are  discrete-time histories in a standard quantum theory (whose
underlying Hilbert space is denoted $\H$),
then there exists an operator $X$ on $(\ot^n\H)\ot(\ot^m \H)$ such that
the decoherence functional may be written as
\beq
	d(\a,\b) = \tr_{(\ot^n \H)\ot(\ot^m \H)}\left(\a\ot\b X \right)
\eeq
where $X$ is independent of
$\a\eqdef\a_{t_1}\ot\a_{t_2}\ot\cdots\ot\a_{t_n}$ and
$\b\eqdef\b_{t_1'}\ot\b_{t_2'}\ot\cdots\ot\b_{t_m'}$.  It was
further shown that the above construction could be extended to the
full infinite tensor product $\V^\Omega
\eqdef\ot^\Omega_{t\in\mathR} \H_t$, the space upon which
discrete-time propositions are projection operators.  Thus the
result $d(\a,\b) = \tr\left(\a\ot\b X \right)$ extends to the set
$\UP$ of all discrete-time history propositions in standard quantum
theory.

	In the same paper we classified the decoherence functionals in the
case when $\UP$ is the lattice of propositions $\P(\V)$ where the
Hilbert space $\V$ has a finite dimension.  In particular, we proved
that the four axioms of section 1, namely {\em hermiticity\/}, {\em
positivity\/}, {\em additivity\/} and {\em normalisation\/}, suffice
to show that every decoherence functional $d$ can be written in the
form
\beq
	d(\a,\b) = \tr_{\V\ot\V} \left(\a\ot\b X \right)
\eeq
for some operator $X$ on the tensor product space $\V\ot\V$.

	The purpose of this section is to show that the decoherence
functional of standard quantum mechanics with {\em continuous\/}
time projections can also be written in the form
\beq
	d(\m,\n) = \tr_{\Vcts\ot\Vcts}\lbra P_\m \ot P_\n X\rbra
\eeq
where $P_\m$ and $P_\n$ are suitable projection operators on
$\Vcts$, and $X$ is an operator on $\Vcts\ot\Vcts$.

	However, before embarking on this discussion we need first to
specify what we mean by a continuous product of projectors in
standard quantum theory, and then calculate $d(\m,\n)$ for these
projectors.  The previous discussion leads one to think that it
might be particularly appropriate to consider continuous products of
projectors onto coherent states. Therefore, let
$\l[a,b]\rightarrow \mathC$ be a curve of bounded
variation, and let us try to define the continuous product of
coherent-state projection operators in the Heisenberg picture.

	We begin with the simple case where the Hamiltonian is zero, so
that there is no difference between Schr\"odinger picture and
Heisenberg picture.  Hence we are interested in
\beq
	\C_{[a,b]}[\l] \eqdef \prod_{t\in [a,b]} P_{\l(t)}	\label{CtsProd}
\eeq
where
\beq
	P_{\l(t)} \eqdef\ketbra{{\l(t)}}{{\l(t)}} 			\label{Projlt}
\eeq
and $\ket{z}$ denotes the usual normalised coherent state.

	In order to give meaning to the right hand side of \eq{CtsProd} let us
start by subdividing the interval $[a,b]$ as $a=t_1<t_2<\ldots <t_n=b$.
A finite approximation to $\C_{[a,b]}[\l]$ is then
\beqa
	&& \mbox{\hskip-5em}P_{\l(t_1)}P_{\l(t_2)}\ldots P_{\l(t_n)}\nn	\\
	&=& \ket{\l(t_1)}
			\braket{{\l(t_1)}}{{\l(t_2)}}\braket{{\l(t_2)}}{{\l(t_3)}}
			\bra{\l(t_3)}\ldots\ketbra{{\l(t_{n})}}{{\l(t_n)}}	\nn	\\
	&=& e^{-\half\lbra |\l(t_1)|^2+ |\l(t_n)|^2\rbra
			-\lbra |\l(t_2)|^2+ |\l(t_3)|^2+\ldots
					|\l(t_{n-1})|^2 \rbra} 						\nn	\\
	&{}&\quad\times\ e^{\l(t_1)^*\l(t_2) +
		\l(t_2)^*\l(t_3)\ldots +\l(t_{n-1})^*\l(t_n)}\times\
			\ketbra{{\l(a)}}{{\l(b)}}							\nn	\\
	&=& e^{\l(t_1)^*\lbra\l(t_2)-\l(t_1)\rbra
		+ \l(t_2)^*\lbra\l(t_3)-\l(t_2)\rbra +\ldots +
			\l(t_{n-1})^*\lbra\l(t_{n})-\l(t_{n-1})\rbra}	\nn	\\
	&{}&\times\ e^{\half\lbra |\l(t_1)|^2- |\l(t_n)|^2\rbra}
		\ \times\ \ketbra{{\l(a)}}{{\l(b)}}.
\eeqa
As the subdivision gets finer and finer, the first exponent on the
right hand side converges to the {\em Stieltjes\/} integral
$\int_a^b\l^*d\l$, which motivates defining the continuous product
\eq{CtsProd} as
\beqa
	\C_{[a,b]}[\l]
		&\eqdef&e^{\int_a^b\l^*d\l}e^{\half\lbra|\l(a)|^2-|\l(b)|^2\rbra}\,
			\times\ \ketbra{{\l(a)}}{{\l(b)}}							\nn	\\
		&=&	e^{{1\over 2}\int_a^b(\l^*\,d\l-\l\,d\l^*)}
				\times\ \ketbra{{\l(a)}}{{\l(b)}}.		\label{CtsProdSt}
\eeqa
Note that the operator thus defined satisfies
\beq
	\C_{[a,b]}[\l]\,\C_{[b,c]}[\l]=\C_{[a,c]}[\l]		\label{C-semigroup}
\eeq
where $a<b<c$. This semigroup property is a natural consistency
condition to impose on any definition of a continuous product of
projection operators.

	We turn now to the case in which the Hamiltonian is non-zero, and
first define the Heisenberg-picture operator, with fiducial time
$t_0$,
\beq
	P_{\l(t)}(t) \eqdef U(t_0,t)P_{\l(t)}U(t,t_0),
\eeq
so that a finite-time approximation to the product $\prod_{t\in [a,b]}
P_{\l(t)}(t)$ is
\beqa
	&& U(t_0,t_1)P_{\l(t_1)}U(t_1,t_2)P_{\l(t_2)}
		U(t_2,t_3)P_{\l(t_3)}\ldots P_{\l(t_n)}  U(t_n,t_0)		\nn	\\
	&=&\ U(t_0,t_1)\ket{\l(t_1)}\bra{\l(t_1)}U(t_1,t_2)\ket{\l(t_2)}
		\bra{\l(t_2)}U(t_2,t_3)\ket{\l(t_3)}\bra{\l(t_3)}\ldots	\nn	\\
	&{}&\quad\bra{\l(t_{n-1})}U(t_{n-1},t_{n})\ket{\l(t_{n})}
					\bra{\l(t_n)}U(t_n,t_0).
\eeqa
We now use a standard trick \cite{Klau85a} to evaluate the matrix
elements for small $(t_r-t_{r+1})$:
\beqa
\lefteqn{\bra{\l(t_r)}U(t_r,t_{r+1})\ket{\l(t_{r+1})} =
\bra{\l(t_r)}e^{-i(t_r-t_{r+1})H/\hbar}\ket{\l(t_{r+1})}}\hspace{2cm}\nn\\
	&&\approx\bra{\l(t_r)}{1-i(t_r-t_{r+1})H/\hbar}\ket{\l(t_{r+1})} \nn\\
	&&\approx\braket{{\l(t_r)}}{{\l(t_{r+1})}}\times				\nn	\\
	&&\qquad\lbra{1-i(t_r-t_{r+1})H\lbra\l(t_r),\l(t_{r+1})
				\rbra/\hbar}\rbra									\nn	\\
	&&\approx\braket{{\l(t_r)}}{{\l(t_{r+1})}}\,
			e^{-i(t_r-t_{r+1})H\lbra\l(t_r),\l(t_{r+1})\rbra/\hbar}
\eeqa
where
\beq
	H(z,z^\prime)\eqdef{\bra{z}H\ket{z^\prime} \over\braket{z}{z^\prime}},
\eeq
and where we are not attempting to be rigorous about domains of
unbounded operators and the like. Thus, in the limit of finer and
finer subdivisions, we again get an expression for the continuous
product of projectors as an integral
\beqa
\C_{[a,b]}[\l] &{=}& \prod_{t\in [a,b]} P_{\l(t)}(t) 	\nn	\\
	&=& e^{\half\lbra |\l(a)|^2 - |\l(b)|^2 \rbra}
		e^{\int_a^b\l^*d\l +i/\hbar \int_a^b H(\l(t))\,dt}	\nn	\\
	&{}&\quad \times U(t_0,a)\ket{\l(a)}\bra{\l(b)}U(b,t_0)
								\label{cont-prod-proj}
\eeqa
where $H(z)\eqdef H(z,z)=\bra{z}H\ket{z}$. Note that this definition
also satisfies the semigroup property \eq{C-semigroup}.

	We can now use these results to compute the decoherence
functional for the histories associated with the paths $t\mapsto
\m(t)$ and $t\mapsto\l(t)$, where from now on we will take $a=
-\infty$ and $b=\infty$ for simplicity.  Thus we define
\beq
	d(\m,\n) \eqdef \tr
\left\{\prod_{t\in\mathR} \left(P_{\m(t)}(t)\right)^\dagger
\r_{-\infty}\prod_{t\in\mathR} \left(P_{\n(t)}(t)\right)\right\}
\eeq
where $\r_{-\infty}$ is the initial density matrix. Using our previous
results we can calculate this decoherence functional as
\beq
	d(\m,\n) = e^{\int_{-\infty}^\infty (\n^*d\n - \m^*d\m)}
			e^{{i/\hbar}\int_{-\infty}^\infty (H(\n(t)) - H(\m(t)))\,dt}
				\bra{0}\r_{-\infty}\ket{0}.	\label{dfinteg}
\eeq
Note that, as might have been expected, there is no non-trivial way that
fine-grained histories of this sort will decohere.

\subsection{The decoherence functional as a trace over the Hilbert
space $\Vcts\ot\Vcts$}
The aim now is to write \eq{dfinteg} in the form
\beq
	d(\m,\n) = \tr_{\Vcts\ot\Vcts} \left\{ P_{\eket\m\cdot}
		\ot P_{\eket\n\cdot }\ X\right\}
\eeq
where $P_{\eket\m\cdot}$ and $P_{\eket\n\cdot}$ are the {\em bona fide\/}
projection operators constructed in section 2.

	It is helpful to construct $X$ in stages.  Firstly, consider the
following trace of \eq{cont-prod-proj} in the case that the dynamics
is trivial (\ie the Hamiltonian is zero)
\beq
	\tr_{L^2(\mathR)}\Big(\prod_{t\in\mathR} P_{\l(t)}\Big) =
			e^{\int_{-\infty}^\infty \l^* d\l}.	\label{trProdP}
\eeq
where we have used the fact that $\l$ is square integrable, and
hence vanishes at $\pm\infty$. We wish to write this as a trace over
$\Vcts=\ot_t L^2_t(\mathR)$ \ie we wish to find an operator $\Scts$ such that
\beq
	\tr_{L^2(\mathR)}\Big(\prod_{t\in\mathR} P_{\l(t)}\Big)=
			\tr_\Vcts\lbra \Scts P_{\eket\l\cdot}\rbra
\eeq
In order to understand how to construct $\Scts$, let us recall first what
happens for a discrete, $n$-time history. As shown in \cite{IL94b},
in this case we can write
\beq
	\tr_{L^2(\mathR)}\lbra A_1 A_2 \ldots A_n\rbra
		= \tr_{\ot^n L^2(\mathR)}\lbra A_1\ot A_2 \ot\ldots\ot A_n S_n\rbra
\eeq
where $S_n$ is the operator on the $n$-fold tensor product space
$\ot^n L^2(\mathR)$ that acts on the vector $v_1\ot v_2 \ot\ldots\ot v_n\ \in\
\ot^n L^2(\mathR)$ as
\beq
	S_n (v_1\ot v_2 \ot\ldots\ot v_n) = v_2\ot v_3 \ot\ldots\ot
			v_n\ot v_1.					\label{Sn_def}
\eeq

	For the purposes of the present paper it is important to note
that $S_n$ is closely related to a discrete version of the
derivative operator $\Delta_n$.  If we take the time step $\d t$ to
have value one, we can write the action of $\Delta_n$ on a function
symbolically as
\beq
	\Delta_n f(x_r) = f(x_{r+1}) - f(x_r)
\eeq
\ie $\Delta_n$ corresponds to the matrix
\beq
	\Delta_n = \lbra \begin{array}{llllll}
								 -1 & 1 &0& & &  \\
								 0 & -1 & 1  \\
								   & & & \ddots \\
								 &&&&&\cdots
						\end{array}\rbra
\eeq
Although $S_n$ acts on the $n$-fold tensor product and not on a direct sum of
Hilbert spaces, one can nevertheless use the same notation to
characterise its action as
\beq
	S_n\lbra\begin{array}{l}
					v_1 	\\
					v_2 	\\
					\vdots 	\\
					\vdots	\\
					v_n
			\end{array}\rbra
	=
 		\lbra\begin{array}{l}
					v_2 	\\
					\vdots 	\\
					\vdots	\\
					v_n		\\
					v_1
			\end{array}\rbra
\eeq
\ie
\beq
	S_n = \lbra\begin{array}{llllll}
					0 & 1 &0& & &  \\
					0 & 0 & 1  \\
					 & & & \ddots \\
					 &&&&&
				\end{array}\rbra
\eeq
In other words, and returning to the continuous case, we might expect that
$\Scts$ is roughly $1+\Delta$.  Therefore, we might try to define
$\Scts$ by
\beq
	\Scts\eket\n\cdot = |\exp(\n(\cdot) + \dot{\n}(\cdot))\rangle.
\eeq
or, perhaps better, define the
operator $\Scts$ by its matrix elements:
\beq
	\ebra \m \cdot\Scts \eket \n \cdot \eqdef
	e^{\inner \m\n + \inner{\m}{\dot{\n}} }			\label{Scts_def}
\eeq
where
\beq
	e^{\inner{\m}{\dot{\n}}} \eqdef e^{\int_{-\infty}^\infty \m^* d\n}
\eeq
in which the right hand side is a Stieltjes integral. Note that
\eq{Scts_def} defines $\Scts$ uniquely since the set of vectors of
the form $\eket\n\cdot$, where $\n$ is differentiable, is total.

	The next step is to calculate the trace of the operator $\Scts
P_{\eket\l\cdot}$ on the Hilbert space $\Vcts$. In the case of simple
non-normalised coherent states $\ket{\exp z}$ on $L^2(\mathR)$, the
trace of any trace-class operator $T$ can be written as
\beq
	\tr(T)=\int d\k(z)\bra{\exp z}T\ket{\exp z} \label{resunit}
\eeq
where
\beq
	d\k(z)={1\over\pi}e^{-|z|^2}d(\Re z)\,d(\Im z)
\eeq
and where $\Re z$ and $\Im z$ denote the real and imaginary parts
respectively of the complex number $z$.

	The important thing for us is the analogue of this expression
for the exponential Hilbert space. The key step here is the
observation that the exponential Hilbert space $\exp L^2(\mathR)$ is
isomorphic to a Hilbert space of functionals on $L^2(\mathR)$ whose
inner product is defined via a certain Gaussian measure $\D\m$. The
details are standard \cite{Guich72} (albeit a little tricky in the
functional analysis sense) and, for our purposes, it suffices to say
that the generating functional for this measure has the property
\beq
	\int \D\m[\n]\,e^{\inner\n\l+\inner\rho\n}=e^{\inner\rho\l}.
\eeq
It can be shown that the trace of a trace-class operator $T$ on the
exponential Hilbert space can be evaluated using this measure as
\beq
		\tr(T)=\int\D\m[\n]\,\bra{\exp\n(\cdot)}T\ket{\exp\n(\cdot)},
\eeq
which is the desired analogue of \eq{resunit}.

	We may thus calculate $\tr_\Vcts\lbra \Scts P_{\eket \l \cdot} \rbra$ as
\beqa
\tr_\Vcts\lbra \Scts P_{\eket \l \cdot} \rbra &=&
\int \D\m[\n]\,\ebra \n \cdot \Scts P_{\eket \l \cdot} \eket \n \cdot \nn\\
&=&
\int \D\m[\n]\,\ebra \n \cdot \Scts \eket \l \cdot e^{\inner \l {\n-\l}}\nn\\
&=&
\int \D\m[\n]\, e^{\inner \n {\l+\dot{\l}} + \inner \l {\n-\l}}\nn\\
&=&
  e^{\inner\l{\l+\dot{\l}} - \inner{\l}{\l}}\nn\\
&=&
  e^{\inner \l {\dot{\l}}}\nn\\
&=&
e^{\int_{-\infty}^\infty \l^* d\l}
\eeqa
which is in agreement with \eq{trProdP}, as desired.

	The next step is to introduce dynamics, which we do by finding
an operator $\U$ such that
\beq
	e^{\inner{\l}{\dot{\l}}}e^{{i\over\hbar}H[\l]} =
		\tr_\Vcts\lbra \Scts\,\U P_{\eket\l\cdot}\rbra	\label{Prop:U}
\eeq
where $H[\l]\eqdef \int_{-\infty}^\infty H(\l(t))\,dt$, and where
we note that, from \eq{cont-prod-proj}, the left hand side of
\eq{Prop:U} is $\bra{0}U(-\infty,\infty)\ket{0}^{-1}
\tr(\prod_{t\in\mathR}P_{\l(t)}(t))$. The trace on the right
hand side of \eq{Prop:U} can be evaluated as
\beqa
&& \mbox{\hskip-5em}\tr_\Vcts\lbra \Scts\,\U P_{\eket \l \cdot }\rbra\ \nn\\
&=&
\int\D\m[\eta]\,\ebra\eta\cdot\Scts\,\U P_{\eket\l\cdot}\eket\eta\cdot\nn\\
&=&
	\int \D\m[\eta]\,\ebra\eta\cdot P_{\eket\l\cdot}
		\Scts\,\U P_{\eket \l \cdot} \eket\eta\cdot					\nn\\
&=&
	\int \D\m[\eta]\,e^{\inner {\eta-\l}{\l}}
		\ebra \l \cdot  \Scts\,\U  \eket\l\cdot e^{\inner {\l}{\eta-\l}}\nn\\
&=&
	\ebra \l \cdot \Scts\,\U\eket{\l}{\cdot}e^{-2\inner {\l}{\l}}
	\int\D\m[\eta]\, e^{\inner\eta\l+\inner{\l}{\eta}} \nn\\
&=&
	\ebra\l\cdot\Scts\,\U\eket{\l}{\cdot}e^{-\inner {\l}{\l}}.
\eeqa
Hence we are able to write the trace of the product of projection operators
$\tr_{L^2(\mathR)}(\prod_{t\in\mathR} P_{\l(t)}(t))$ as
$\bra{0}U(-\infty,\infty)\ket{0}\,\tr_{\Vcts}(\Scts\,\U
P_{\eket\l\cdot})$ using the operator $\Scts\,\U$ defined by
\beq
	\ebra\l\cdot\Scts\,\U\eket{\l}{\cdot}
	=	e^{\inner{\l}{\l +\dot{\l}}}e^{{i\over\hbar}H[\l]}
	= \ebra\l\cdot\Scts\,\eket{\l}{\cdot}e^{{i\over\hbar}H[\l]}\label{SctsUdef}
\eeq
While we shall not need it explicitly in what follows, it is not
difficult to show that the matrix elements of $\U$ are given by
\beqa
\ebra{\eta}{\cdot}\U\eket{\n}{\cdot}
	&=&e^{\lbra\inner{\lbra 1+{\partial\over\partial t}\rbra^{-1\dagger}\eta}
		{{\d\over\d\bar{\l} }} + \inner{{\d\over\d\l}}{\m}\rbra}\nn\\
	&{}&\quad \times e^{\inner{\l}{\l +\dot{\l}}}
		e^{{i\over\hbar}H[\l]}\ \big|_{\l=\bar{\l}=0}
\eeqa
where we have used the fact that, for coherent states, the off-diagonal
matrix elements of any operator $A$ are given in terms of the diagonal ones
by
\beq
	\ebra{\eta}{\cdot} A \eket{\n}{\cdot}
		= e^{\lbra\inner\eta{{\d\over\d\bar{\l}}}
			+ \inner{{\d\over\d\l}}{\n} \rbra}\\
  \ebra\l\cdot A \eket{\l}\cdot \ \big|_{\l=\bar{\l}=0}
\eeq

	We are now in a position to express the decoherence functional in terms
of a trace over $\Vcts\ot\Vcts$.  We recall that, from \eq{dfinteg},
\beqa
d(\m,\n)
	&=&\tr_{L^2(\mathR)}\bigg(\Big(\prod_{t\in\mathR}P_{\m(t)}\Big)^\dagger
		\r_{-\infty}\Big(\prod_{t\in\mathR}P_{\n(t)}\Big)\bigg) 		\nn\\
	&=& e^{\int_{-\infty}^\infty (\n^*d\n - \m^*d\m)}
			e^{{i/\hbar}\int_{-\infty}^\infty (H(\n(t)) - H(\m(t)))\,dt}
				\bra{0}\r_{-\infty}\ket{0}.
\eeqa
Thus, using the operator $\Scts\,\U$ defined by \eq{SctsUdef}, we get
\beqa
d(\m,\n)
 &=& \tr_\Vcts\lbra \Scts\,\U P_{\eket\m\cdot}\rbra^*
      \tr_\Vcts\lbra \Scts\,\U P_{\eket\n\cdot}\rbra
      \bra{0}\r_{-\infty}\ket{0}\nn\\
&=&   \tr_\Vcts\lbra \lbra\Scts\,\U\rbra^\dagger P_{\eket\m\cdot}\rbra
      \tr_\Vcts\lbra \Scts\,\U P_{\eket\n\cdot}\rbra
      \bra{0}\r_{-\infty}\ket{0}\nn\\
&=&   \tr_{\Vcts\otimes\Vcts}\lbra P_{\eket\m\cdot}\ot P_{\eket\n\cdot}X\rbra
\eeqa
where
\beq
	X\eqdef\bra{0}\r_{-\infty}\ket{0}\,
		\lbra\Scts\,\U\rbra^\dagger\ot\lbra\Scts\,\U\rbra.
\eeq

\bigskip\noindent
\section{Conclusion}
We have shown how the history group of a continuous-time history
theory leads to a certain natural class of projection operators in a
continuous tensor-product space. We have also shown how to construct
class operators as continuous products of projection operators, and
we demonstrated how the decoherence function of a pair of class
operators can be re-expressed in the tensor product space using the
projection operators that represent the history propositions.  The
calculations of decoherence functionals performed above are
restricted to histories associated with projectors onto coherent
states: an important task for the future therefore is to extend
these calculations to include projectors that correspond to temporal
averages of physical quantities.

	The history group plays a central role in these constructions by
giving us the Hilbert space $\Vcts$ whose projectors form the space
of history propositions, just as the canonical group gives the
Hilbert space of standard, single-time quantum mechanics. The fact
that the history group of quantum mechanics is the same as the
canonical group of a $1+1$-dimensional {\em field\/} theory raises
several important issues. The first is the existence of many
unitarily inequivalent representations: a feature whose relevance
for the history theory remains to be seen. Another is the intriguing
possibility that arises in the case of a spin system.  The
single-time canonical commutation relations are
\beq
	\[S^i,S^j\] = i\e^{ijk} S^k, \quad i,j,k = 1,2,3 \label{Spin}
\eeq
which suggests that the relevant history group has the Lie algebra
\beq
	\[S^i_t,S^j_{t'}\] = i\e^{ijk} S^k_{t'}\d(t-t'),
							\quad i,j,k = 1,2,3. 	\label{CtsSpin}
\eeq
However, unlike \eq{Spin}, the algebra \eq{CtsSpin} can have a non-trivial
central extension, which raises the possibility that the history algebra for a
spin system might instead be
\beq
	\[S^i_t,S^j_{t'}\] = i\e^{ijk} S^k_{t'} \d(t-t')
			+ i\k\d^{ij}\partial_t\d(t-t') \label{CtsSpinx}
\eeq
where $\k$ is a constant. The representation theories of \eq{CtsSpin}
and \eq{CtsSpinx} are, of course, very different.  The investigation of
the physical implications of using \eq{CtsSpinx} rather than \eq{CtsSpin}
will be left to future work.

	There is another feature of the history framework that does not
occur in the standard single-time treatment, and which deserves
further study: the fact that there are `temporally entangled' states
in the history theory.  While this feature appears in the case of
continuous histories studied in this paper, it also
occurs in a theory with only discrete-time histories, and is perhaps
simpler to describe there.

	Consider the case of standard quantum mechanics, and let us consider only
two-time histories.  If the Hilbert space of the canonical theory is
$\V$, then the Hilbert space upon which the history propositions are
projection operators is $\V\ot\V$.  As well as the homogeneous histories,
(projectors onto vectors of the form $\ket{u}\ot\ket{v}$) and the
inhomogeneous histories made from them by the appropriate logical
operations (e.g. disjoint union $\oplus$), the space of projectors on
$\V\ot\V$ also includes certain `exotic histories', for
example the projector onto the suitably normalised vector
\beq
\lbra \ket{u_1}\ot\ket{v_1} + \ket{u_2}\ot\ket{v_2} \rbra
\eeq
In the case of the single-time quantum theory of a composite system,
this is the sort of entangled state that leads to many of the peculiar
features of quantum mechanics. In the case of the history framework, we
see that there is the possibility of a quite new type of
quantum entanglement.

\bigskip
\noindent{\large\bf Acknowledgements}

\noindent
We would like to thank Jim Hartle for many helpful discussions
during early stages of the work. We gratefully acknowledge support
from the SERC, from Trinity College, Cambridge (CJI), and the
Leverhulme and Newton Trusts (NL).  The work described in this paper
was started at the Isaac Newton Institute, and we are both grateful
to the staff there for providing a stimulating working environment.

\end{document}